\documentclass[twocolumn]{aastex63}
\usepackage{color,graphicx,amsmath,cases}

\newcommand{\vrad}{{$v_{\mathrm{r}}$}~}

\newcommand{\kms}{{km s$^{-1}$}~}
\newcommand{\kmsend}{{km s$^{-1}$}}

\submitjournal{the AAS Journals}

\shorttitle{Stellar Population Gradients in the Virgo Core GC System}
\shortauthors{Ko et al.}

\begin{document}

\title{The Next Generation Virgo Cluster Survey. XXXIII. \\ Stellar Population Gradients in the Virgo Cluster Core Globular Cluster System}

\correspondingauthor{Youkyung Ko, Eric Peng}
\email{ykko@kasi.re.kr, peng@pku.edu.cn}

\author{Youkyung Ko}
\affiliation{Korea Astronomy and Space Science Institute, 776 Daedeok-daero, Yuseong-Gu, Daejeon 34055, Korea}
\affiliation{Department of Astronomy, Peking University, Beijing 100871, China}
\affiliation{Kavli Institute for Astronomy and Astrophysics, Peking University, Beijing 100871, China}

\author{Eric W.\ Peng}
\affiliation{Department of Astronomy, Peking University, Beijing 100871, China}
\affiliation{Kavli Institute for Astronomy and Astrophysics, Peking University, Beijing 100871, China}

\author{Patrick C{\^o}t{\'e}}
\affiliation{Herzberg Astronomy and Astrophysics Research Centre, National Research Council of Canada, 5071 W. Saanich Road, Victoria, BC, V9E 2E7, Canada}

\author{Laura Ferrarese}
\affiliation{Herzberg Astronomy and Astrophysics Research Centre, National Research Council of Canada, 5071 W. Saanich Road, Victoria, BC, V9E 2E7, Canada}

\author{Chengze Liu}
\affiliation{Department of Astronomy, School of Physics and Astronomy, and Shanghai Key Laboratory for Particle Physics and Cosmology, Shanghai Jiao Tong University,
Shanghai 200240, People's Republic of China}

\author{Alessia Longobardi}
\affiliation{Dipartimento di Fisica G. Occhialini, Universit{\'a} degli Studi di Milano Bicocca, Piazza della Scienza 3, I-20126 Milano, Italy}

\author{Ariane Lan{\c{c}}on}
\affiliation{Universit{\'e} de Strasbourg, CNRS, Observatoire astronomique de Strasbourg, UMR7550, F-67000 Strasbourg, France}

\author{Roberto P.\ Mu{\~n}oz}
\affiliation{Institute of Astrophysics, Pontificia Universidad Cat{\'o}lica de Chile, Av. Vicu{\~n}a Mackenna 4860, 7820436 Macul, Santiago, Chile}

\author{Thomas H.\ Puzia}
\affiliation{Institute of Astrophysics, Pontificia Universidad Cat{\'o}lica de Chile, Av. Vicu{\~n}a Mackenna 4860, 7820436 Macul, Santiago, Chile}

\author{Karla A. Alamo-Mart{\'i}nez}
\affiliation{Department of Astronomy, DCNE, Universidad de Guanajuato, Callejon de Jalisco s/n, 36023, Guanajuato, Mexico}

\author{Laura V. Sales}
\affiliation{University of California Riverside, 900 University Ave., Riverside CA 92521, USA}

\author{Felipe Ramos-Almendares}
\affiliation{CONICET - Universidad Nacional de C{\'o}rdoba, Instituto de Astronom{\'i}a Te{\'o}rica y Experimental (IATE), Laprida 854, X5000BGR, C{\'o}rdoba, Argentina}
\affiliation{Instituto de Investigaci{\'o}n Multidisciplinar en Ciencia y Tecnolog{\'i}a, Universidad de La Serena, Ra{\'u}l Bitr{\'a}n 1305, La Serena, Chile}
\affiliation{Departamento de F{\'i}sica y Astronom{\'i}a, Universidad de La Serena, Av. Cisternas 1200 N, La Serena, Chile}

\author{Mario G.\ Abadi}
\affiliation{CONICET - Universidad Nacional de C{\'o}rdoba, Instituto de Astronom{\'i}a Te{\'o}rica y Experimental (IATE), Laprida 854, X5000BGR, C{\'o}rdoba, Argentina}
\affiliation{Observatorio Astron{\'o}mico, Universidad Nacional de C{\'o}rdoba, Laprida 854, X5000BGR, C{\'o}rdoba, Argentina}

\author{Myung Gyoon Lee}
\affiliation{Astronomy Program, Department of Physics and Astronomy, Seoul National University, Gwanak-gu, Seoul 151-742, Republic of Korea}

\author{Ho Seong Hwang}
\affiliation{Astronomy Program, Department of Physics and Astronomy, Seoul National University, Gwanak-gu, Seoul 151-742, Republic of Korea}
\affiliation{SNU Astronomy Research Center, Seoul National University, 1 Gwanak-ro, Gwanak-gu, Seoul 08826, Republic of Korea}
\affiliation{Korea Astronomy and Space Science Institute, 776 Daedeok-daero, Yuseong-Gu, Daejeon 34055, Korea}

\author{Nelson Caldwell}
\affiliation{Center for Astrophysics, Harvard \& Smithsonian, 60 Garden Street, Cambridge, MA 02138, USA}


\author{John P.\ Blakeslee}
\affiliation{NSF’s NOIRLab, 950 N. Cherry Avenue, Tucson, AZ 85719, USA}

\author{Alessandro Boselli}
\affiliation{Aix Marseille Univ, CNRS, CNES, LAM, Marseille, France}

\author{Jean-Charles Cuillandre}
\affiliation{AIM, CEA, CNRS, Universit{\'e} Paris-Saclay, Universit{\'e} Paris Diderot, Sorbonne Paris Cit{\'e}, Observatoire de Paris, PSL University, F-91191 Gif-sur-Yvette Cedex, France}

\author{Pierre-Alain Duc}
\affiliation{Universit{\'e} de Strasbourg, CNRS, Observatoire astronomique de Strasbourg, UMR7550, F-67000 Strasbourg, France}

\author{Susana Eyheramendy}
\affiliation{Faculty of Sciences, engineering and technology, Universidad Mayor, Manuel montt 367, Santiago, Chile}

\author{Puragra Guhathakurta}
\affiliation{Univ. of California Santa Cruz, UCO/Lick Obs., 1156 High St., Santa Cruz, CA 95064, USA}

\author{Stephen Gwyn}
\affiliation{Herzberg Astronomy and Astrophysics Research Centre, National Research Council of Canada, 5071 W. Saanich Road, Victoria, BC, V9E 2E7, Canada}

\author{Andr{\'e}s Jord{\'a}n}
\affiliation{Facultad de Ingenier{\'i}a y Ciencias, Universidad Adolfo Ib{\'a}{\~n}ez, Av. Diagonal las Torres 2640, Pe{\~n}alol{\'e}n, Santiago, Chile}
\affiliation{Millennium Institute of Astrophysics, Santiago, Chile}

\author{Sungsoon Lim}
\affiliation{Department of Astronomy, Yonsei University, 50 Yonsei-ro, Seodaemun-gu, Seoul 03722, Republic of Korea}

\author{Rub{\'e}n S{\'a}nchez-Janssen}
\affiliation{UK Astronomy Technology Centre, Royal Observatory, Blackford Hill, Edinburgh, EH9 3HJ, UK}

\author{Elisa Toloba}
\affiliation{Department of Physics, University of the Pacific, 3601 Pacific Avenue, Stockton, CA 95211, USA}




\begin{abstract}
We present a study of the stellar populations of globular clusters (GCs) in the Virgo Cluster core with a homogeneous spectroscopic catalog of 692 GCs within a major axis distance $R_{\rm maj} = $ 840~kpc from M87. We investigate radial and azimuthal variations in the mean age, total metallicity, [Fe/H], and $\alpha$-element abundance, of blue (metal-poor) and red (metal-rich) GCs using their co-added spectra. We find that the blue GCs have a steep radial gradient in [Z/H] within $R_{\rm maj} =$ 165~kpc, with roughly equal contributions from [Fe/H] and [$\alpha$/Fe], and flat gradients beyond. By contrast, the red GCs show a much shallower gradient in [Z/H], which is entirely driven by [Fe/H]. We use GC-tagged Illustris simulations to demonstrate an accretion scenario where more massive satellites (with more metal- and $\alpha$-rich GCs) sink further into the central galaxy than less massive ones, and where the gradient flattening occurs because of the low GC occupation fraction of low-mass dwarfs disrupted at larger distances. The dense environment around M87 may also cause the steep [$\alpha$/Fe] gradient of the blue GCs, mirroring what is seen in the dwarf galaxy population. The progenitors of red GCs have a narrower mass range than those of blue GCs, which makes their gradients shallower. We also explore spatial inhomogeneity in GC abundances, finding that the red GCs to the northwest of M87 are slightly more metal-rich. Future observations of GC stellar population gradients will be useful diagnostics of halo merger histories.
\end{abstract}





\section{Introduction \label{section:intro}}

The assembly of the most massive galaxies in the Universe is a story of continual merging and accretion of gas, stars, and dark matter over a Hubble time. These behemoths have long ago stopped forming significant numbers of their own stars {\it in situ}, but are still adding stars to their halos through the accretion of lower mass galaxies \citep{ose10, jno12}. When massive galaxies are situated at the centers of galaxy clusters, their stellar halos (``cD envelopes") join seamlessly into the larger accreted population sometimes referred to as ``intracluster light" \citep{mur04, mur07, pbz07, con14}. While the relatively low surface brightnesses of these halo stars make them difficult to study, they contain a wealth of information on the progenitor components that assembled to form the parent galaxy. In the most massive galaxies, the accreted halo stars may actually be the dominant stellar population \citep{pbz07, ose10, coo13}.

The characterization of extragalactic halo stellar populations is made challenging by their low surface brightnesses, usually well below that of the sky. Deep imaging surveys of massive galaxies have revealed extended stellar halos in cluster galaxies \citep{ber95, gon00, fel04, mih05, kb07, rud10, mih17}, some showing significant fine structure expected from merging \citep{duc15, bil20}, and some have been resolved into individual stars \citep{mou05a, mou05b, mou07, wil12}. Obtaining further information on the nature of these stellar halos is more difficult, as even colors are subject to uncertainties with sky subtraction \citep[e.g.,][]{mih13}. Recent spectroscopic surveys have been able to derive the integrated ages, metallicities, and $\alpha$-element abundances of the brighter regions ($R <$ 3 effective radii) of the stellar halos of some massive early-type galaxies \citep[e.g.,][]{spo10, gre15, gu18a, gu18b, iod19, fen20}.

Numerous studies have used discrete tracers like planetary nebulae \citep[PNe; e.g.][]{hui93, arn96, arn98, cia04, bac06, cia10, cor13, lon13, lon15a} and globular clusters \citep[GCs; e.g.][]{bea00, pen06, hwa08, liu11, ush15, pas15, lim17, pow18, lon18a, ko18, ko19, ko20} with success to elucidate the stellar content of halo populations.
In particular, colors of old GCs are indicative of metallicities, which have been broadly used to trace the mass assembly history of host galaxies. The GCs in massive galaxies commonly show bimodal color distributions, with the blue and red GCs corresponding to metal-poor and metal-rich sub-populations, respectively. The properties of red GCs show tighter correlations with the underlying stars in their host galaxies in terms of their spatial distributions \citep[e.g.][]{pl13,wan13} and kinematics \citep[e.g.][]{pen04,pot13}. Several studies have suggested a scenario where the red (metal-rich) GCs were formed with the bulk of the stars during dissipative mergers, while the blue (metal-poor) GCs were formed in small galaxies and have been continuously accreted onto massive galaxies \citep[e.g.][]{cot98, lee10}.

We can better understand these possibilities by combining chemical and spatial information. Metallicity gradients, a common feature of galaxies, inform us about the role of the gravitational potential in the regulation of star formation and chemical enrichment, but also highlight the role of major mergers in erasing these gradients \citep[e.g.,][]{spo10}. Gas-poor minor merging can also create metallicity gradients, as higher mass galaxies with more metal-rich stars have a tendency to sink further into the host galaxy due to dynamical friction \citep{amo19}. For example, in the Milky Way's stellar halo, the existence or not of abundance gradients has been important in shaping our view of how the halo may have formed. The lack of a metallicity gradient for Milky Way GCs beyond 8 kpc led \citet{sz78} to propose that the outer halo was formed from the accretion of low-mass fragments.

The color gradients of GC systems have been explored in a relatively large sample of early-type galaxies in the Virgo and Fornax clusters \citep{liu11}, finding that both the blue (metal-poor) and red (metal-rich) GC systems exhibit radial color gradients, showing equal strengths within the uncertainties. This result holds even when gradient measurements are extended to $R \sim 5$--$8~R_{\rm e}$, as in the literature compilation of \citet{fr18}, but flattening of the gradients of metal-poor GCs was found at $R\gtrsim 8R_{\rm e}$.

Another chemical parameter for understanding the formation history of galaxies is the ratio between $\alpha$-element abundances and iron, [$\alpha$/Fe].
This provides information about the star formation timescale, i.e., a high [$\alpha$/Fe] value indicates rapid star formation and early quenching.
As an example, the correlation of [$\alpha$/Fe] in Milky Way stars with their locations in the Galaxy has been used to understand the formation of thin and thick disks \citep[e.g.][]{rec14}.
For GCs in massive early-type galaxies, it is known that they have a wide range of [$\alpha$/Fe] values and often have super-solar [$\alpha$/Fe] ratios \citep[e.g.,][]{puz05,par12}, which indicates that they were formed over a short timescale. Studies of the spatial variation of [$\alpha$/Fe] in extragalactic GCs, however, are rare, but have the potential to reveal in detail how galaxy halos are assembled.

Recently, there have been attempts to understand the characteristics of the GC populations in the context of cosmological simulations. The E-MOSAICS project \citep{pfe18,kru19,rei19} combined the MOSAICS semi-analytic models for star cluster formation and evolution with the EAGLE hydrodynamical simulation for galaxy formation, including 25 Milky Way-like disk galaxies. They studied the formation and co-evolution of galaxies and their star clusters in terms of chemical abundances, progenitors, accretion epochs of the GCs.
In another approach, \citet{ram20} investigated the GC populations in nine simulated Virgo-like galaxy clusters. They tagged the GCs to the simulated galaxies at their infall time, and traced the kinematics and spatial distributions of the GCs. Other groups have also started to tag GCs in cosmological simulations to study the spatially resolved properties of GC systems \citep{tru21,che22}. These developments have made it possible use the spatial gradients and distributions of GCs in meaningful comparisons to simulations.

The Virgo Cluster is an excellent target to investigate the mass assembly history of the central galaxy because of its proximity. It is a dynamically young system that shows many substructures of member galaxies both spatially and kinematically \citep{bst85,bts87}. In particular, the core region of the Virgo shows diffuse stellar features \citep{mih05,mih17} and extended structures of ionized gas filaments \citep{ken08,bos18} and HI gas \citep{yos02,ov05} around galaxies, indicative of strong galaxy interactions. The most massive galaxy at the center of the Virgo, M87, has a weak cD envelope, and kinematic studies of GCs and PNe have been done to understand the connection between its faint stellar halo and the intracluster light \citep{lon15a,ko17,lon18a,lon18b}. In addition, the GCs and PNe in M87 also constitute kinematic substructures related to recent mergers \citep{str11,rom12,lon15b,oe16,lrv20}. Therefore, M87 and the Virgo cluster are a good laboratory to study the ongoing assembly of the most massive galaxy in a cluster core. 

In addition to the many aforementioned GC kinematic studies, the chemistry and ages of GCs have also been studied from photometry and spectroscopy in Virgo ellipticals \citep[e.g.,][]{cbr98,cbc03}. \citet{pow18} reported a substructure of relatively young GCs located $\sim$ 11$\farcm$5 south of M87, possibly related to past mergers or accretions. More recently, \citet{vil20} extended the study of chemical abundances in GCs out to $\sim$ 140 kpc from M87, showing the promise of stellar population studies in extragalactic GCs across a range of radius. They do not find any significant iron abundance gradients of both metal-poor and metal-rich populations within $R < 40$ kpc and conclude that the flat metallicity gradient of metal-poor GCs could be a sign that some metal-poor GCs were formed {\it in situ} at high-redshift. Moreover, they found the metal-poor GC population has a higher $\alpha$-element abundance than the metal-rich GC population in both the inner ($R < 40$ kpc) and outer ($R > 40$ kpc) halos, which is not expected by current cosmological simulations.

In this study, we investigate mean trends of the stellar populations of the GCs in the core region of the Virgo Cluster out to $\sim$ 840 kpc ($2\fdg9$), as a function of major axis distance from M87, using a new homogeneous spectroscopic catalog. This survey mainly covers a $3\arcdeg \times 2\arcdeg$ field including the massive elliptical galaxies M87, M86 and M84. This paper is organized as follows. The GC catalog used in this study and the sample selection are described in Section \ref{section:data}. In Section \ref{section:method}, we explain how we measure ages and metallicities of the GC sub-populations. We present the main results about the age and metallicity distributions of the GCs in Section \ref{section:results}. We discuss the assembly history of M87 and the Virgo Cluster core in Section \ref{section:discussion}. In Section \ref{section:summary}, we summarize the results. We adopt a distance to the Virgo cluster of 16.5 Mpc \citep{mei07,bla09}, so one arcminute corresponds to 4.91 kpc.


\section{Data and GC Sample Selection \label{section:data}}

	\subsection{Sample Definition}
	
We produced a new homogeneous catalog of the GCs in the Virgo core region by combining two spectroscopic data sets.
The first data set was part of a spectroscopic follow-up program to the Next Generation Virgo Cluster Survey \citep[NGVS;][]{fer12}.
The NGVS is an extensive deep imaging survey using the MegaCam imager on the Canada-France-Hawaii Telescope (CFHT). It images a 104 deg$^2$ field of the Virgo Cluster with five optical filters $u^*g'r'i'z'$. The images were taken with excellent seeing conditions ($< 0\farcs6$ for $i$-band imaging), and many GCs in the Virgo Cluster are marginally resolved in the NGVS images. The $i'$-band ``fuzziness'' index ($\Delta i_{4-8}$), defined as the difference between 4-pixel and 8-pixel $i'$-band aperture-corrected magnitude, has larger values for the GCs than for foreground stars on average \citep{dur14}.
In addition, a $K_s$-band imaging survey was carried out with WIRCam on the CFHT \citep[NGVS-IR;][]{mun14} for a $2\arcdeg \times 2\arcdeg$ region around M87. \citet{mun14} showed that the $u^*i'K_s$ color-color diagram allows us to clearly distinguish Virgo GCs from contaminants such as foreground stars and background galaxies.
We identified GCs in the Virgo core based on a ``fuzziness'' index and various $u^*g'r'i'z'K_s$ color combinations of point sources, adopting the ``extreme deconvolution" (xD) classification algorithm \citep{bov11}. GCs were selected using xD, but also using a broader set of color and morphology cuts, and targeted for spectroscopy. This program obtained spectra of 776 GCs in the Virgo Cluster core, some of which have been used in previous studies \citep{zhu14,li20}.

Second, we used spectra taken from \citet{ko17} who presented a spectroscopic catalog of the GCs in a central $2\arcdeg \times 2\arcdeg$ field of the Virgo Cluster, covering the vicinity of M87, M84 and M86. They obtained 910 optical spectra of GC candidates. We reclassified their sample using the $u^*i'K_s$ color-color diagram for the objects in the NGVS-IR field, and confirmed 212 GCs in their sample.
There are 88 sources in common between the NGVS follow-up program and \citet{ko17}. Finally, we combined these two samples to construct a combined spectroscopic catalog of 903 GCs in total in the Virgo core.

Both programs were taken with the MMT/Hectospec \citep{fab05} with a 270~mm$^{-1}$ grating having a dispersion of 1.2~\rm{\AA}~pixel$^{-1}$. Data were obtained in the wavelength range 3650 -- 9200~\rm{\AA}.
Data reduction was done with the SAO pipeline, following the description in \citet{cal11}, and flux calibration was done following the steps in \citet{fab08}. We used the xcsao task in the IRAF RVSAO package \citep{km98} to measure heliocentric radial velocities of targets, taking 10 different templates including stars, GCs, and galaxies. Combining radial velocity, color, and size information, we classify the objects into four groups: GCs, UCDs, foreground stars, and background galaxies. The details of the data reduction, flux calibration, heliocentric radial velocity measurements, and target classification will be described in a forthcoming paper about the spectroscopic catalog of the GCs (Ko et al. in preparation). The median signal-to-noise ratio ($S/N$) of the spectra of GC candidates at 5000~\rm{\AA} is $S/N \sim 9$, where the Poisson noise, sky noise, and readout noise are considered to calculate spectral noise as a function of wavelength.

We select 726 out of the 903 total GCs, excluding the GCs that belong to other Virgo galaxies in the core. The remaining GCs are either GCs that belong to M87 or intracluster GCs (IGCs), a population governed by the galaxy cluster gravitational potential \citep{wil07, fir08, lph10, dur14}. We did this because our interest for this paper is mainly to investigate the assembly history of M87 in the cluster environment.
There are 154 Virgo galaxies that have radial velocity information in the survey region.
We used both spatial and kinematic criteria to separate the GCs bound to Virgo galaxies, following \citet{lon18a}: 1) $R < 10 R_{\rm e}$ where $R_{\rm e}$ is the effective radius of a given galaxy and 2) $|v_{\rm r}-v_{\rm gal}| < 3\sigma_{\rm gal}$ where \vrad, $v_{\rm gal}$, and $\sigma_{\rm gal}$ are the radial velocities of GCs, galaxy systemic velocity, and galaxy central velocity dispersion, respectively. The radial velocities of galaxies are compiled from various literature sources \citep{ned19, ahn14, kim14, fer20}. The velocity dispersions are from GOLDMine database \citep{gav03} compiling data from \citet{mce95} and \citet{sco98}. For the galaxies where velocity dispersions are not known, mostly faint dwarf galaxies, we adopted $\sigma_{\rm gal} = 50$ \kmsend.

We expect that these cuts are very conservative, and remove from our sample nearly all the GCs that belong to other galaxies. As an example, for larger galaxies, the effective radius of the GC system ($R_{\rm e,GC}$) may be as big as $\sim5$~times that of the stars \citep[e.g.,][]{kar16,cas19}, so our cut to eliminate GCs within $10 R_{\rm e,stars}$ corresponds to a cut of $2 R_{\rm e,GC}$, which results in $\sim85\%$ of GCs being removed, leaving $\sim15\%$ to be potential contaminants. The velocity cut, however, removes all GCs within $\pm3\times$ the velocity dispersion from the mean galaxy velocity. If the GC line-of-sight velocity distribution (LOSVD) is roughly Gaussian, then this selection removes all but $0.27\%$ of the GCs. Combined with the spatial cut, only 0.04\% of the GC system could potentially contaminate our sample. Our spectroscopic limit ($g'<22.5$~mag) and a Gaussian GC luminosity function shape \citep[$\langle g\rangle\approx 24$~mag, $\sigma\sim1.3$~mag;][]{jor07} means that we only observe the brightest $\sim 12\%$ of the GCs in any massive galaxy. Therefore, the fraction of GCs that could enter our sample is roughly $5\times10^{-5}$. The massive nearby galaxies M86 and M84 have $\approx3000$~GCs each \citep{pen08}. The analysis above, therefore, leads us to expect that only 0.3 GCs from these galaxies could contaminate our sample. Of course, this analysis assumes that the GC spatial distribution follows a S{\'e}rsic profile at large radii, and that the LOSVD is Gaussian in the wings. These simple assumptions may not hold in practice, but any GC that is significantly detached (spatially and kinematically) from its host is probably on its way to being accreted by the M87 system, and arguably should be included in our analysis.

\begin{figure*}
\centering
\includegraphics[width=\textwidth]{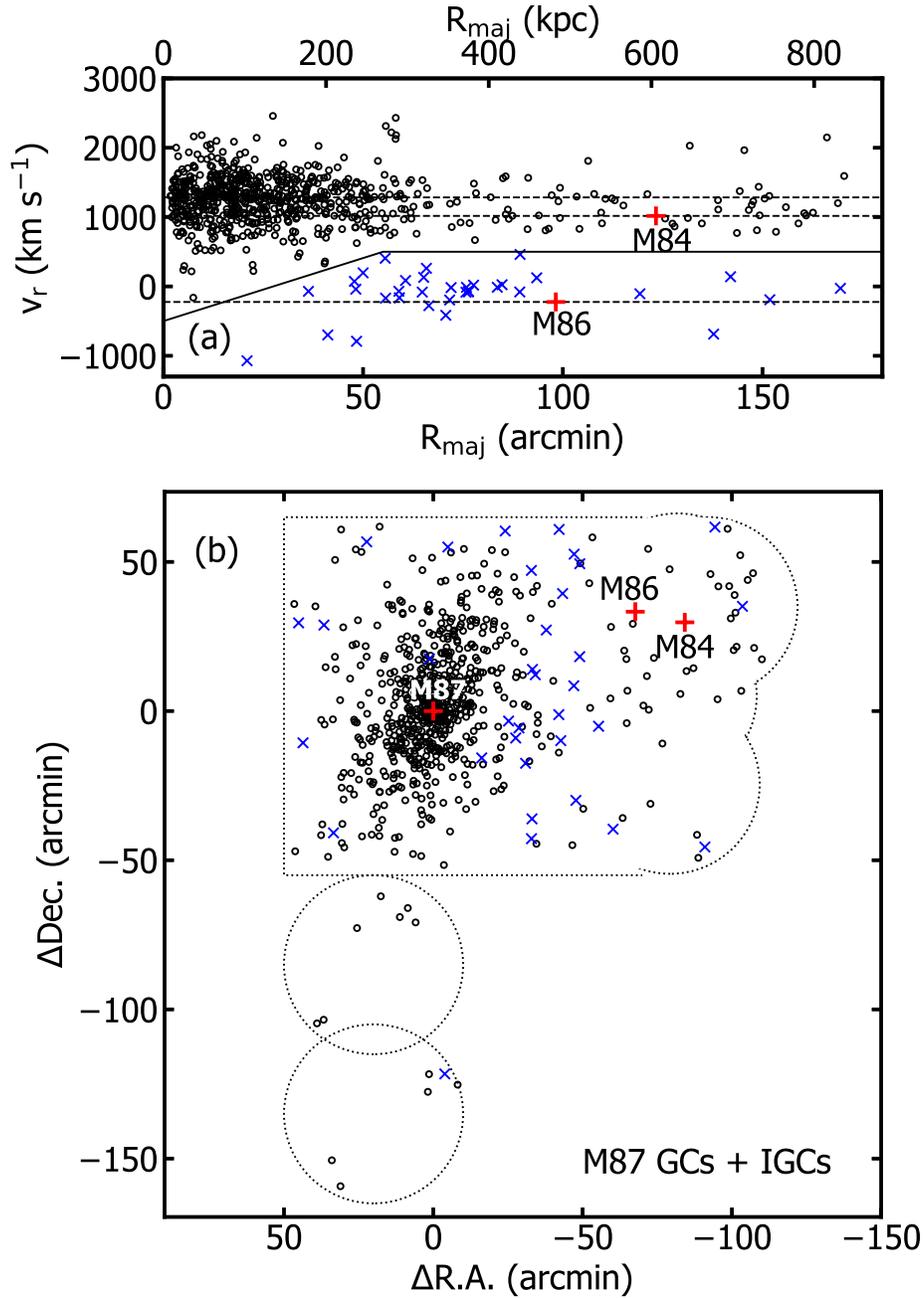}
\caption{(a) Radial velocity vs. major axis distance from M87 for the M87 GCs and IGCs.
The black circles and blue crosses represent the GCs in the Virgo main body and the low velocity substructure of the GCs, respectively. The red pluses indicate the location of M86 and M84.
The systemic velocities of M87 (1284 \kmsend), M86 (--224 \kmsend), and M84 (1017 \kmsend) are marked with horizontal dashed lines \citep{cap11}.
The black solid line indicates the criterion for separating out the low velocity objects.
(b) Spatial distribution of the M87 GCs and IGCs. The center of M87 is at (0, 0), and the symbols are the same as (a).
The dotted outlines show the survey region of this study. \label{fig:subpop}}
\end{figure*}

\begin{figure*}
\centering
\includegraphics[width=0.9\textwidth]{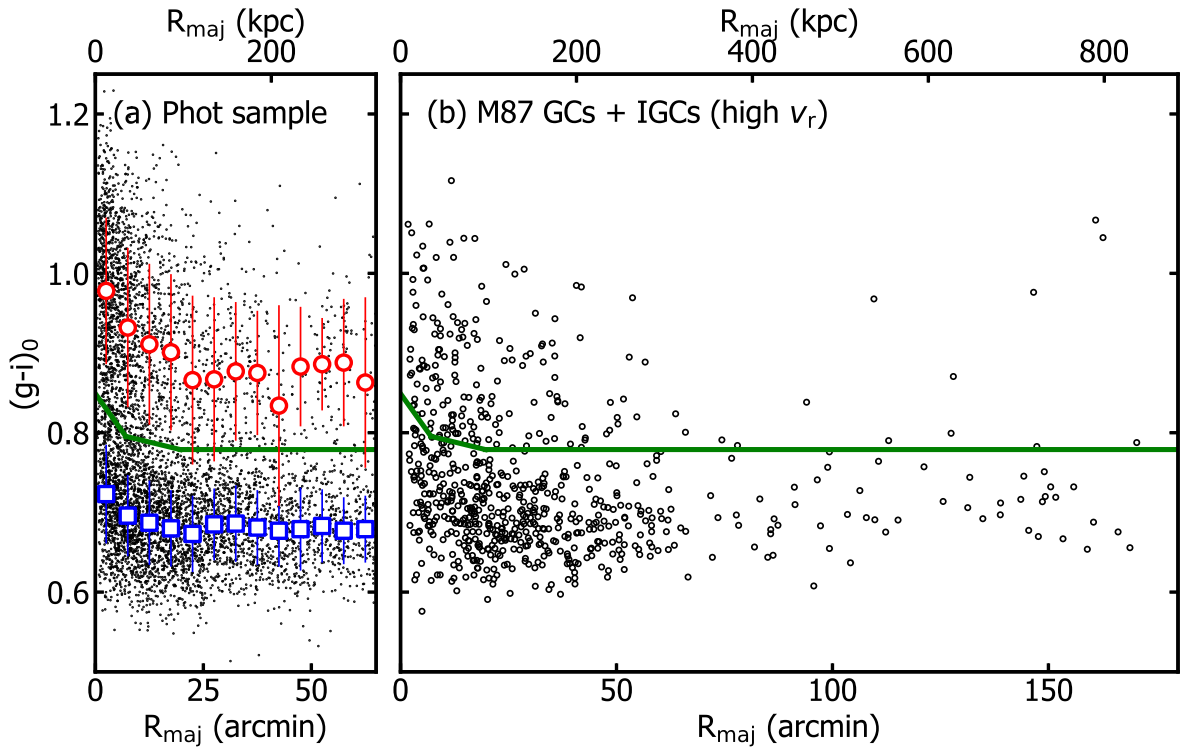}
\caption{(a) $(g'-i')_0$ color vs. major axis distance from M87 for the GC candidates within a $2\arcdeg \times 2\arcdeg$ region around M87, identified in the NGVS photometric survey. The blue squares and red circles represent the mean $(g'-i')_0$ colors of the blue and red GC candidates within each radial bin, respectively, estimated from the GMM. The error bars indicate the Gaussian widths of the two components. The green solid lines indicates the color criterion for dividing the GCs into blue and red GCs. (b) Same as (a), but for the GCs without the low velocity GCs, spectroscopically confirmed in this study. \label{fig:color}}
\end{figure*}

Figure \ref{fig:subpop}(a) shows the radial velocities of 726 GCs as a function of major axis distance from M87. The major axis distance $R_{\rm maj}$ is calculated as follows:
\begin{align*} 
& R_{\rm maj} = [(x/(1-e))^2+y^2]^{1/2} \tag{1} \\
{\rm where}~ x & =\Delta{\rm R.A.}~{\rm cos}({\rm PA}) - \Delta{\rm Dec}~{\rm sin}({\rm PA}) \\
{\rm and}~ y & =\Delta{\rm R.A.}~{\rm sin}({\rm PA}) + \Delta{\rm Dec}~{\rm cos}({\rm PA})
\end{align*}
\noindent

\noindent
with a position angle (PA) of 155$\arcdeg$ and an ellipticity $e$ of 0.4 for M87 \citep{fer06,jan10}. Most GCs have radial velocities consistent with a mean value of 1255 \kmsend, similar to the systemic velocity of M87 \citep[1284 \kmsend;][]{cap11}. In addition, we identify another group of GCs at $30\arcmin < R_{\rm maj} < 150\arcmin$, that appears to be a broad feature in the velocity-position phase-space diagram connecting M87 with M86. The mean radial velocity of these GCs is \vrad $= -66$ \kms, which is about 1370 \kms lower than the systemic velocity of M87. 

Figure \ref{fig:subpop}(b) shows the spatial distribution of the GCs in our study. The GCs are strongly concentrated around M87, although some of them are uniformly distributed in the survey region beyond $R_{\rm maj} = 65\arcmin$. The GCs with low radial velocities are mostly located in the region west of M87 (between it and M86). Note that the low velocity GCs that are strongly concentrated on M86 (systemic velocity of $-224$ \kmsend) are considered to belong to it and are not plotted in this figure. For this analysis, we only used the 692 GCs that have higher radial velocities in order to focus on the main structure of the Virgo Cluster, assuming that the low velocity GCs constitute an infalling group associated with M86.

\begin{figure*}
\includegraphics[width=\textwidth]{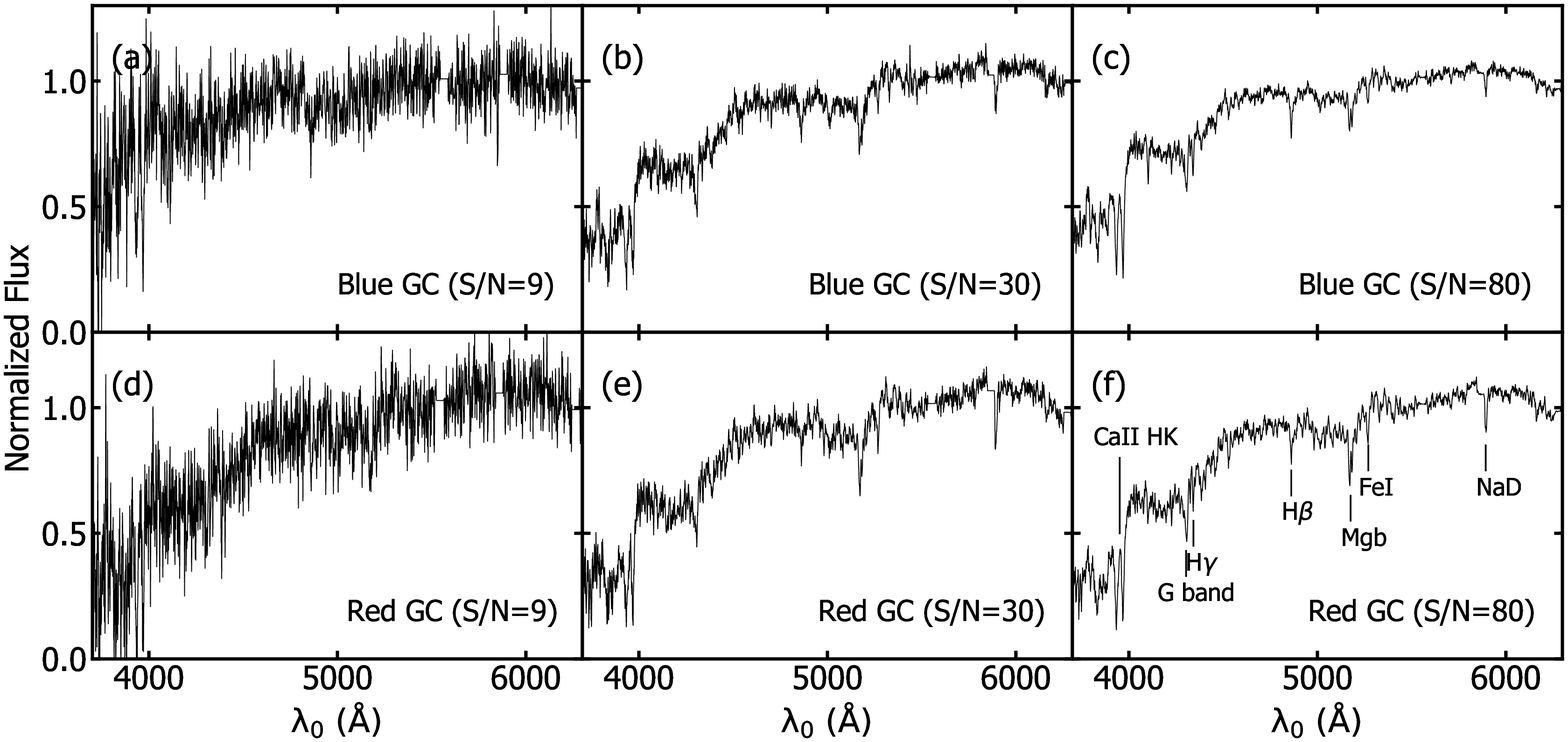}
\caption{Examples of flux-calibrated spectra of (a) a blue GC and (d) a red GC that have $S/N \sim 9$.
Middle and right panels show co-added spectra of GCs with the $S/N$ of $\sim$ 30 and 80, respectively.
The spectra were stacked for (b) three blue GCs at $2\arcmin < R_{\rm maj} < 2\farcm4$, (c) 53 blue GCs at $2\arcmin < R_{\rm maj} < 7\farcm1$, (e) two red GCs at $1\farcm6 < R_{\rm maj} < 2\farcm2$, and (f) 45 red GCs at $1\farcm6 < R_{\rm maj} < 9\farcm2$.
All spectra were normalized at 5450--5550 \AA. \label{fig:sample_spec}}
	\end{figure*}

\subsection{GC Subsamples}

Many studies have confirmed the existence of color bimodality for GCs in M87 \citep{whi95,es96,kun99,lar01,tam06,pen06,pen09,har09}, a feature typical of GCs in massive early-type galaxies. Since we expect that the blue and red GC sub-populations have different formation origins, we divided our GC sample into blue and red, using their $(g'-i')_0$ colors, to search for any differences in their stellar populations.

We performed Gaussian Mixture Modeling \citep{mg10} on the $(g'-i')_0$ color distribution of the GCs in a $2\arcdeg \times 2\arcdeg$ region around M87, identified in the NGVS photometric survey (see Section 2.1) within radial bins five arcminutes in width, assuming heteroscedastic bimodal distributions. Figure \ref{fig:color}(a) shows the $(g'-i')_0$ color of the photometric sample as a function of major axis distance from M87.
We adopted the color at which the probability that GCs have a 50\% probability of belonging to either component as the $(g'-i')_0$ color that divides the two components.
This color criterion changes as a function of major axis distance. We adopted a broken linear fit for the dividing color for GCs with $R_{\rm maj} < 20\arcmin$ as follows:

\begin{subequations}
	\begin{align}
\begin{split}
    (g'-i')_0 = {}& 0.8490 - 0.0760 \times R_{\rm maj}, \\
         & \mbox{if $R_{\rm maj} < 7\arcmin$,}
\end{split}\\
\begin{split}
    (g'-i')_0 = {}& 0.8048 - 0.0013 \times R_{\rm maj}, \\
         & \mbox{if $7\arcmin < R_{\rm maj} < 20\arcmin$.}
\end{split}
	\end{align}
\end{subequations}

\noindent
The color criterion becomes almost flat at $(g'-i')_0 = 0.779$ beyond $R_{\rm maj} = 20\arcmin$. It is worth mentioning that the main radial gradient results are not dependent on the details of the color criterion, and are little changed even if we adopt a single color at all radii, such as $(g-i)_0 = 0.8$~mag.

Figure \ref{fig:color}(b) shows the $(g'-i')_0$ color of our GC samples as a function of major axis distance from M87. The majority of the GCs are located within $R_{\rm maj} = 65\arcmin$ where a strong sequence at $(g'-i')_0 \sim 0.7$ and a weak red tail are shown. Most of the red GCs are located within $R_{\rm maj} = 40\arcmin$, while the blue GCs extend out to $R_{\rm maj} = 180\arcmin$.
We have 509 blue and 183 red GCs in our sample with the color criteria described above.


\section{Age, [Z/H], [F\lowercase{e}/H], and [\texorpdfstring{$\alpha$}{a}/F\lowercase{e}] Measurements \label{section:method}}

The colors of GCs are an efficient tool to measure their metallicities. However, there are inherent uncertainties because GC colors also depend on GC ages. A spectral analysis has the advantage of measuring ages and metallicities of GCs more independently. 

Lick line indices have been commonly used to measure ages and metallicities of old simple stellar populations with their integrated spectra.
They are defined as the strengths of 25 prominent absorption lines seen in optical spectra of old stellar systems, established and refined by \citet{bur84}, \citet{wor94}, \citet{wo97}, and \citet{tra98}.

We used the lick\_ew code in the EZ\_Ages package \citep{gs08} to measure the Lick indices from our spectra, adopting the passbands defined by \citet{wo97} and \citet{tra98}.
This code smooths spectra to the resolution of the Lick system and estimates index errors based on error spectra \citep{car98}.

We measured the mean ages, metallicities, and $\alpha$-element abundances of the GC subgroups using a $\chi^2$ minimization technique \citep{pro04}.
This technique is based on the residuals between the Lick index measurements and the prediction from stellar population models.
We used the flux-calibrated stellar population models of Lick indices presented by \citet{tmj11}.
The models were constructed with the ages ranging from 0.1 to 15 Gyr, the metallicities [Z/H] from $-2.25$ to $+0.67$, and the $\alpha$-element abundances, [$\alpha$/Fe], from $-0.3$ to $+0.5$. We measured the iron abundances by adopting the relation from \citet{tmb03}, [Z/H] = [Fe/H] + 0.94 [$\alpha$/Fe]. 

We used all Lick indices except CN$_1$, CN$_2$, Ca4227, and NaD to calculate the $\chi^2$ values.
The stellar population model we adopted is calibrated with Galactic GCs that show an anomaly in the nitrogen abundance, which might not be applicable to typical simple stellar populations.
The CN$_1$, CN$_2$, and Ca4227 are sensitive to nitrogen abundances so they were excluded from the fitting.
The NaD index is also excluded from the fitting because it is strongly affected by interstellar absorption.
We determined the $\chi^2$ values with the remaining Lick indices, using iterative 2$\sigma$ clipping.
Finally, 14 to 21 line indices of each co-added spectrum are used for the fitting.

Because our individual GC spectra typically do not have the required signal-to-noise ratio ($S/N$) for a robust stellar population analysis, we performed our analysis on co-added spectra. We stacked the spectra of various GC subgroups to obtain mean ages, metallicities, and $\alpha$-element abundances by taking the error weighted mean values for each wavelength bin after strong sky emission line regions were masked.
All spectra were de-redshifted prior to the co-addition.
Figure \ref{fig:sample_spec}(a) and (d) show flux-calibrated spectra of typical blue and red GC spectra, with $S/N\sim 9$ per \AA\ at 5000\AA, which is the median $S/N$ for our sample. Panels (b), (c), (e), and (f) show stacked spectra of blue and red GCs with $S/N$ of 30 and 80 as examples. The stacking analysis enables us to determine stellar population parameters from high $S/N$ spectra for GC subgroups. We have made the stacked spectra used in our analysis publicly available on Zenodo (\href{https://doi.org/10.5281/zenodo.6390374}{doi:10.5281/zenodo.6390374})\footnote{These data will be available on Zenodo when the paper is published in the journal. Those interested in accessing the data before publication are welcome to contact the corresponding authors.}.

For the following analysis, the measurement errors for the ages, [Z/H], [Fe/H], and [$\alpha$/Fe] were calculated using the bootstrap.
For a given GC subgroup, we chose the same number of GCs in the group allowing replacement, stacked their spectra, and measured the stellar population parameters.
We repeated this process 1000 times, and took the 16th and 84th percentiles of the parameter distributions for lower and upper errors.
The following results show the stellar population analysis for the GCs stacked using various criteria.


\begin{figure*}
\centering
\hspace{-0.2cm}\includegraphics[width=0.8\textwidth]{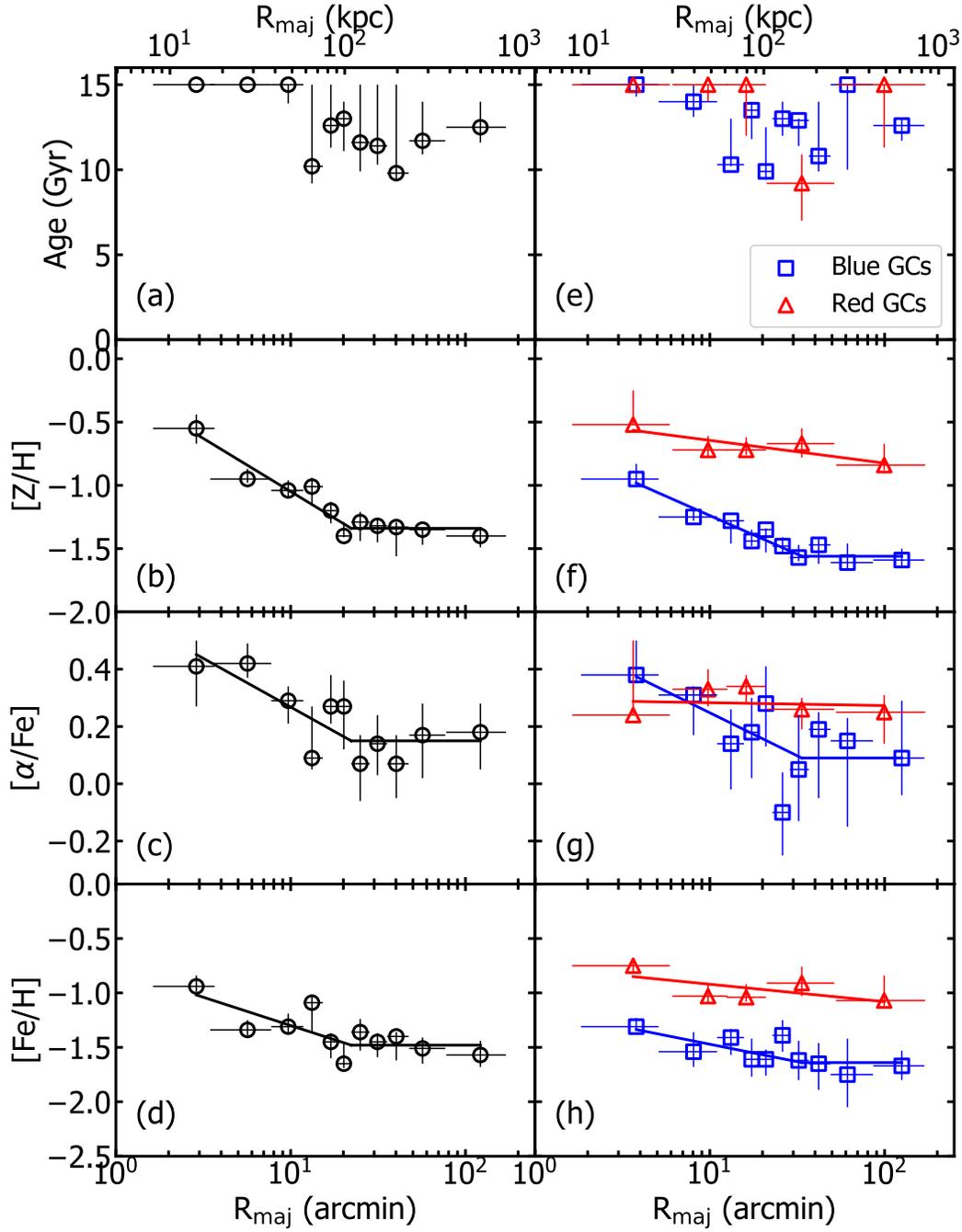}
\caption{(a) Mean age, (b) mean [Z/H], (c) mean [$\alpha$/Fe], and mean [Fe/H] of the GC population as a function of major axis distance from M87.
Panels (e)-(h) show the same as panels (a)-(d), respectively, but for GC sub-populations: blue GCs (blue squares) and red GCs (red triangles).
The solid lines in panels (b)-(d) and (f)-(h) represent the broken linear or linear fits for the GCs, blue and red GCs. The error bars on the y-axis represent the 1-sigma uncertainties from bootstrapping, while the errors bars in $R_{\rm maj}$ along the x-axis represent the radial extent of each bin. \label{fig:r_measure}}
	\end{figure*}

\section{Results \label{section:results}}

\begin{deluxetable*}{c c c c c c}
\tablecaption{Mean ages, [Z/H], [$\alpha$/Fe], and [Fe/H] for the entire GC sample \label{tab:rsp_gc}}
\tablehead{
\colhead{$R_{\rm maj}$ range (kpc)} & \colhead{$\overline{R_{\rm maj}}^a$ (kpc)} & \colhead{Age (Gyr)} & \colhead{[Z/H] (dex)} & \colhead{[$\alpha$/Fe] (dex)} & \colhead{[Fe/H] (dex)}
}
\startdata
$8 - 18$ & $14$ & $15.0^{+<0.1}_{-<0.1}$ & $-0.55^{+0.11}_{-0.12}$ & $0.41^{+0.09}_{-0.14}$ & $-0.94^{+0.10}_{-0.05}$ \\
$17 - 38$ & $28$ & $15.0^{+<0.1}_{-<0.1}$ &  $-0.95^{+0.08}_{-0.03}$ & $0.42^{+0.07}_{-0.05}$ & $-1.34^{+0.09}_{-0.05}$ \\
$38 - 58$ & $48$ & $15.0^{+<0.1}_{-1.1}$ & $-1.04^{+0.08}_{-0.05}$ & $0.29^{+0.05}_{-0.08}$ & $-1.31^{+0.12}_{-0.07}$ \\
$58 - 75$ & $65$ & $10.2^{+4.8}_{-1.0}$ & $-1.01^{+0.07}_{-0.14}$ & $0.09^{+0.18}_{-0.04}$ & $-1.09^{+0.07}_{-0.28}$ \\
$75 - 90$ & $83$ & $12.6^{+2.4}_{-1.3}$ & $-1.20^{+0.07}_{-0.10}$ & $0.27^{+0.11}_{-0.06}$ & $-1.45^{+0.08}_{-0.15}$ \\
$90 - 108$ & $99$ & $13.0^{+1.0}_{-1.9}$ & $-1.40^{+0.11}_{-0.01}$ & $0.27^{+0.09}_{-0.15}$ & $-1.65^{+0.20}_{-0.04}$ \\
$109 - 139$ & $126$ & $11.6^{+3.4}_{-1.7}$ & $-1.29^{+0.08}_{-0.15}$ & $0.07^{+0.10}_{-0.13}$ & $-1.36^{+0.12}_{-0.17}$ \\
$139 - 175$ & $154$ & $11.4^{+3.6}_{-1.1}$ & $-1.32^{+0.05}_{-0.13}$ & $0.14^{+0.10}_{-0.11}$ & $-1.45^{+0.08}_{-0.14}$ \\
$175 - 232$ & $197$ & $9.8^{+5.2}_{-0.4}$ & $-1.33^{+0.04}_{-0.23}$ & $0.07^{+0.10}_{-0.12}$ & $-1.40^{+0.04}_{-0.22}$ \\
$234 - 376$ & $279$ & $11.7^{+2.3}_{-0.8}$ & $-1.35^{+0.01}_{-0.12}$ & $0.17^{+0.11}_{-0.15}$ & $-1.51^{+0.10}_{-0.14}$ \\
$381 - 837$ & $598$ & $12.5^{+1.5}_{-0.9}$ & $-1.40^{+0.06}_{-0.09}$ & $0.18^{+0.10}_{-0.13}$ & $-1.57^{+0.13}_{-0.11}$ \\
\enddata
\tablenotetext{a}{Luminosity-weighted mean major axis distance of GCs in each radial bin.}
\end{deluxetable*}

\begin{deluxetable*}{c c c c c c}
\tablecaption{Mean ages, [Z/H], [$\alpha$/Fe], and [Fe/H] for the blue GC sample \label{tab:rsp_bgc}}
\tablehead{
\colhead{$R_{\rm maj}$ range (kpc)} & \colhead{$\overline{R_{\rm maj}}^a$ (kpc)} & \colhead{Age (Gyr)} & \colhead{[Z/H] (dex)} & \colhead{[$\alpha$/Fe] (dex)} & \colhead{[Fe/H] (dex)}
}
\startdata
$9 - 25$ & $19$ & $15.0^{+<0.1}_{-0.7}$ & $-0.95^{+0.12}_{-0.04}$ & $0.38^{+0.12}_{-0.03}$ & $-1.31^{+0.08}_{-0.03}$ \\
$25 - 54$ & $40$ & $14.0^{+1.0}_{-0.9}$ & $-1.25^{+0.07}_{-0.03}$ & $0.31^{+0.03}_{-0.14}$ & $-1.54^{+0.18}_{-0.14}$ \\
$54 - 77$ & $65$ & $10.3^{+2.7}_{-0.1}$ & $-1.28^{+<0.01}_{-0.18}$ & $0.14^{+0.12}_{-0.16}$ & $-1.41^{+0.08}_{-0.16}$ \\
$77 - 94$ & $85$ & $13.5^{+0.5}_{-1.7}$ & $-1.44^{+0.09}_{-0.04}$ & $0.18^{+0.12}_{-0.16}$ & $-1.61^{+0.19}_{-0.16}$ \\
$94 - 112$ & $103$ & $9.9^{+2.6}_{-0.5}$ & $-1.35^{+0.01}_{-0.18}$ & $0.28^{+0.13}_{-0.15}$ & $-1.61^{+0.09}_{-0.15}$ \\
$112 - 142$ & $128$ & $13.0^{+1.0}_{-1.0}$ & $-1.48^{+0.07}_{-0.05}$ & $-0.10^{+0.14}_{-0.15}$ & $-1.39^{+0.14}_{-0.15}$ \\
$141 - 181$ & $158$ & $12.9^{+0.1}_{-1.5}$ & $-1.57^{+0.10}_{-0.03}$ & $0.05^{+0.19}_{-0.18}$ & $-1.62^{+0.18}_{-0.18}$ \\
$181 - 241$ & $206$ & $10.8^{+3.2}_{-0.9}$ & $-1.47^{+0.05}_{-0.15}$ & $0.19^{+0.06}_{-0.24}$ & $-1.65^{+0.19}_{-0.24}$ \\
$241 - 421$ & $301$ & $15.0^{+<0.1}_{-5.0}$ & $-1.61^{+0.15}_{-0.05}$ & $0.15^{+0.08}_{-0.30}$ & $-1.75^{+0.33}_{-0.30}$ \\
$423 - 829$ & $615$ & $12.6^{+0.4}_{-0.9}$ & $-1.59^{+0.09}_{-0.06}$ & $0.09^{+0.20}_{-0.13}$ & $-1.67^{+0.14}_{-0.13}$ \\
\enddata
\tablenotetext{a}{Luminosity-weighted mean major axis distance of blue GCs in each radial bin.}
\end{deluxetable*}

\begin{deluxetable*}{c c c c c c}
\tablecaption{Mean ages, [Z/H], [$\alpha$/Fe], and [Fe/H] for the red GC sample \label{tab:rsp_rgc}}
\tablehead{
\colhead{$R_{\rm maj}$ range (kpc)} & \colhead{$\overline{R_{\rm maj}}^a$ (kpc)} & \colhead{Age (Gyr)} & \colhead{[Z/H] (dex)} & \colhead{[$\alpha$/Fe] (dex)} & \colhead{[Fe/H] (dex)}
}
\startdata
$8 - 29$ & $18$ & $15.0^{+<0.1}_{-<0.1}$ & $-0.52^{+0.27}_{-<0.01}$ & $0.24^{+0.26}_{-<0.01}$ & $-0.75^{+0.05}_{-0.06}$ \\
$30 - 62$ & $48$ & $15.0^{+<0.1}_{-1.0}$ & $-0.72^{+0.11}_{-0.05}$ & $0.33^{+0.07}_{-0.06}$ & $-1.03^{+0.11}_{-0.06}$ \\
$61 - 103$ & $79$ & $15.0^{+<0.1}_{-3.0}$ & $-0.72^{+0.10}_{-0.04}$ & $0.34^{+0.04}_{-0.06}$ & $-1.04^{+0.12}_{-0.04}$ \\
$104 - 253$ & $165$ & $9.2^{+1.7}_{-2.2}$ & $-0.67^{+0.12}_{-0.11}$ & $0.26^{+0.04}_{-0.07}$ & $-0.91^{+0.15}_{-0.12}$ \\
$259 - 837$ & $488$ & $15.0^{+<0.1}_{-3.7}$ & $-0.84^{+0.17}_{-0.04}$ & $0.25^{+0.06}_{-0.11}$ & $-1.07^{+0.23}_{-0.05}$ \\
\enddata
\tablenotetext{a}{Luminosity-weighted mean major axis distance of red GCs in each radial bin.}
\end{deluxetable*}

\begin{deluxetable*}{l c c c c c c c c c}
\tablecaption{Linear fit results for radial gradients of mean [Z/H], [$\alpha$/Fe], and [Fe/H] for the GC subsample.\label{tab:fit}}
\tablehead{
\colhead{Sample} & \colhead{$R_{\rm maj}$ range (kpc)} & \colhead{$R_0$ (kpc)} & \colhead{$a_\mathrm{[Z/H]}$} & \colhead{$b_\mathrm{[Z/H]}$} & \colhead{$a_\mathrm{[\alpha/Fe]}$} & \colhead{$b_\mathrm{[\alpha/Fe]}$} & \colhead{$a_\mathrm{[Fe/H]}$} & \colhead{$b_\mathrm{[Fe/H]}$}
}
\startdata
All & $8 - R_0$ & $109^{+82}_{-2}$ & $-1.34^{+<0.01}_{-0.12}$ & $-0.84^{+0.15}_{-0.11}$ & $0.15^{+0.04}_{-0.04}$ & $-0.34^{+0.11}_{-0.11}$ & $-1.48^{+0.03}_{-0.07}$ & $-0.52^{+0.06}_{-0.14}$ \\
& $R_0 - 837$ & $109^{+82}_{-2}$ & $-1.34^{+<0.01}_{-0.12}$ & 0 & $0.15^{+0.04}_{-0.04}$ & 0 & $-1.48^{+0.03}_{-0.07}$ & 0 \\
Blue & $9 - R_0$ & $165^{+32}_{-60}$ & $-1.56^{+0.05}_{-0.05}$ & $-0.61^{+0.05}_{-0.21}$ & $0.09^{+0.04}_{-0.10}$ & $-0.30^{+0.08}_{-0.17}$ & $-1.64^{+0.09}_{-0.05}$ & $-0.32^{+0.16}_{-0.10}$ \\
& $R_0 - 829$ & $165^{+32}_{-60}$ & $-1.56^{+0.05}_{-0.05}$ & 0 & $0.09^{+0.04}_{-0.10}$ & 0 & $-1.64^{+0.09}_{-0.05}$ & 0 \\
Red & $8 - 837$ & $100^a$ & $-0.70^{+0.09}_{-0.01}$ & $-0.18^{+0.09}_{-0.15}$ & $0.28^{+0.04}_{-0.04}$ & $-0.01^{+0.05}_{-0.18}$ & $-0.97^{+0.08}_{-0.02}$ & $-0.16^{+0.15}_{-0.05}$ \\
\enddata
\tablecomments{The functional form of linear fits is $a + b~\times$ log($R_{\rm maj}/R_{0}$).}
\tablenotetext{a}{For red GCs, $R_0$ is not a break point of segmented linear fits, but simply a reference point.}
	\end{deluxetable*}
	
	\subsection{Radial Gradients \label{section:rgrad}}
	
Figure \ref{fig:r_measure}(a)-(d) show the mean age, [Z/H], [$\alpha$/Fe], and [Fe/H] of the GC population in each radial bin as a function of major axis distance from M87. The measurement values are listed in Table \ref{tab:rsp_gc}. 
The co-added spectrum of the GCs in each radial bin has a $S/N$ of 80.
Note that the numbers of blue and red GCs in each radial bin are comparable within $R = 14\arcmin$, but the blue GCs dominate in the outer region where its number fraction rises up to $\sim$ 80\%.

The GC populations in all radial bins are older than 8~Gyr, but the ones in the outskirts have formal mean ages about 2~Gyr younger than those in the inner region. However, it is well known that changing horizontal branch morphology can affect age estimates of integrated stellar populations \citep[e.g.,][]{lee00,kol08,cab22}, especially for metal-poor GCs. Tests with our integrated spectra show that age estimates using only H$\delta$ (and masking all other Balmer lines) are 1--2 Gyr younger than those using H$\beta$ alone for the GC bins at large radii, which is a sign that blue horizontal branch stars may be having a moderate effect on age estimates \citep{sch04}. Therefore, although we report this apparent gradient in age, we urge some caution in their interpretation. 

We find a steep radial gradient in the mean metallicities of the GCs within a certain radius, and a flat gradient beyond. We derive broken linear fits for these data with the form as follows:
\begin{align*} 
{\rm [Z/H]} & = a + b \times {\rm log}\left(\frac{R_{\rm maj}}{R_0}\right),&{\rm if}~R_{\rm maj} < R_0, \\
 & = a,&{\rm if}~R_{\rm maj} \geq R_0. \tag{2}
\end{align*}
\noindent
The coefficient $a$ is the value of [Z/H] at and beyond the scale distance $R_0$ (in kpc), and the coefficient $b$ is the slope within $R_0$. The best-fit values we estimated are listed in Table \ref{tab:fit}. 
The entire GC sample shows a steep metallicity gradient with a slope of $b = -0.84^{+0.15}_{-0.11}$ within $R_{\rm maj} \sim 22 \arcmin$ (109 kpc) and a constant metallicity of [Z/H] = $-1.34$ beyond. This value of $b$ means that, within $R_0$, [Z/H] changes by $-0.84$~dex for a factor of ten in radius. The break point of $R_{\rm maj} \sim$ 109 kpc is consistent with the distance at which the radial velocity dispersion of the GC system rises, found in \citep{lon18a}, where the IGC population becomes dominant. 

The inner radial bins have estimated ages that saturate at the limit of the models (15~Gyr). This model age zeropoint issue could potentially affect our metallicity measurements. To test this, we artificially limit the model ages to be below 10~Gyr, finding that the estimated [Z/H] increases by only $\sim0.1$~dex. This moderate increase is likely representative of the potential effects of age saturation, and is much smaller than the magnitude of the detected gradient.

We also confirmed that the metallicities of individual GCs with $S/N > 20$ are derived down to [Z/H] $\sim -2$ with uncertainties of $\sim$ 0.1. In addition, integrated metallicities of the Milky Way GCs, measured using \citet{sch05} spectra with the same method, are distributed below [Z/H] = $-1.6$ and show a good agreement with the values from the literature. We therefore conclude that the flattening of the metallicity gradient in Figure \ref{fig:r_measure}(b) is real, and not due to a lack of metallicity sensitivity.

The $\alpha$-element abundances also show a decreasing trend out to $R_{\rm maj} \sim 22\arcmin$ and then become constant at larger distances. We fitted the data with the same functional form as used for [Z/H] (Eq.\ 2), with $R_0$ fixed to the value derived for [Z/H]. The coefficients $a$ and $b$ are derived to be $a = 0.15 \pm 0.04$ and $b = -0.34 \pm 0.11$. Described in words, this value of $b$ reflects the fact that GCs in M87's inner region are roughly twice as abundant in alpha-elements relative to iron compared to GCs at and beyond $R_0$. We perform a similar analysis for the radial trend of [Fe/H]. Here we also find a significant gradient within $R_0$. Although we again fixed $R_0$ to 109~kpc, allowing $R_0$ to vary freely produces a very similar value. The coefficients $a$ and $b$ for [Fe/H] are estimated to be $a = -1.48^{+0.03}_{-0.07}$ and $b = -0.52^{+0.06}_{-0.14}$. The magnitude of the slopes for [$\alpha$/Fe] and [Fe/H] are comparable within the uncertainties, showing that they each contribute roughly equally to the [Z/H] gradient. 


We also investigate the radial trends of the stellar populations of GC sub-populations where the co-added spectrum in each radial bin has a $S/N$ of 70. The mean ages, [Z/H], [$\alpha$/Fe], and [Fe/H] for blue and red GC samples are listed in Tables \ref{tab:rsp_bgc} and \ref{tab:rsp_rgc}, respectively.
The blue and red GCs in most radial bins have mean ages older than 10 Gyr (Figure \ref{fig:r_measure}(e)).
Figure \ref{fig:r_measure}(f) shows the metallicity gradients of the blue and red GCs.
The red GCs are clearly more metal-rich than the blue GCs, which indicates that the $(g'-i')$ color corresponds well to the average metallicity.

We fit linear relations for the metallicity and $\alpha$-element abundance gradients of the blue sub-population in the same way as they were derived for the entire sample. For the red GCs, we do a single linear fit, as there is no convincing evidence of a break point. The best-fit coefficients are listed in Table \ref{tab:fit}.
The blue GCs have a steep total metallicity gradient with a slope of $b = -0.61^{+0.05}_{-0.21}$ within the $R_{\rm maj} \sim 34\arcmin$ (165 kpc) and a constant metallicity of [Z/H] = $-1.56$ beyond, which is a similar trend to that of the entire sample.
On the other hand, the red GCs have much shallower gradient in [Z/H], with a slope of $b = -0.18^{+0.09}_{-0.15}$. A value of ${\rm [Z/H]} \approx -0.7$ describes the red GCs across a large range in radius.

Figure \ref{fig:r_measure}(g) shows the radial trends of $\alpha$-element-to-iron abundance ratios of the blue and red GCs. The blue GCs show a noticeable gradient with a slope of $b = -0.30^{+0.08}_{-0.17}$ within $R_{\rm maj} \sim 34\arcmin$ (165 kpc), while the red GCs have a constant [$\alpha$/Fe] value of $\sim$ 0.3 in all radial ranges with a radial gradient slope of $b = -0.01^{+0.05}_{-0.18}$. The mean [$\alpha$/Fe] value of the outermost blue GCs is $+0.09$, similar to the solar abundance.

Recently, \citet{vil20} presented a stellar population study of M87 GCs based on full spectrum fitting of individual GC spectra. They found no significant radial gradient in [Fe/H] for either blue or red GCs. A direct comparison to this work is complicated by the fact that the sample of \citet{vil20} covers a smaller range in radius ($R<150$~kpc), and a gap in their data at 50~kpc that motivated their choice of break point in their fitting. Our [Fe/H] gradient for the blue GCs ($-0.32$~dex per decade in radius) is consistent with their findings, within the uncertainties, although a more homogenous analysis between the two datasets is certainly desirable in the future. \citet{vil20} also find a strong negative gradient in the blue GCs for certain $\alpha$-elements, like [Mg/Fe], which is similar to what we see in this work.


\begin{figure}
\hspace{-1cm}\includegraphics[width=1.3\columnwidth]{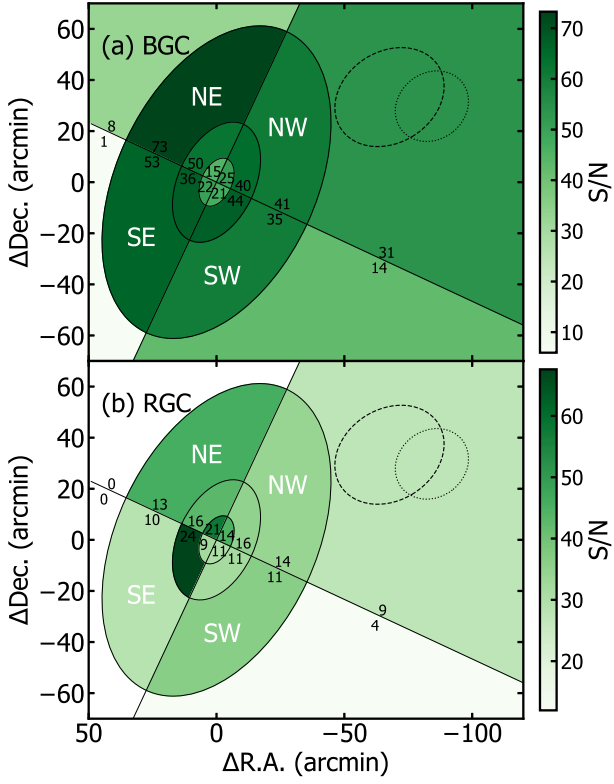}
\caption{(a) $S/N$ of the co-added spectra for the blue GC population in spatial bins.
The solid-line ellipses are centered on M87, and have semi-major axes of 10$\arcmin$, 25$\arcmin$ and 65$\arcmin$, a position angle of 155$\arcdeg$, and an ellipticity of 0.4.
The dashed-line and dotted-line ellipses are centered on M86 and M84 with semi-major axes of 23$\arcmin$ and 15$\arcmin$, corresponding to 10 effective radii of M86 and M84, respectively.
The position angle and ellipticity of M86 are 123$\arcdeg$ and 0.21, and those of M84 are 128$\arcdeg$ and 0.14 \citep{fer20}.
The radial solid lines indicate the position angle criteria used to divide the survey region into different sectors. The numbers in each subregion indicate the numbers of the co-added GCs.
(b) Same as (a), but for the red GC population.
\label{fig:map_sn}}
\end{figure}

\subsection{Azimuthal Variations \label{section:azimuthal}}

The Virgo cluster core around M87 is known to be a place of active galaxy and cluster assembly that is likely not well-mixed \citep[e.g.,][]{mih17}. Our wide spatial coverage of GCs in this region allows us to investigate the possibility that there may also be azimuthal variations of mean ages and total metallicities of the GC population in addition to radial trends. 
We take ellipses centered on M87 that have a position angle of 155$\arcdeg$ and an ellipticity of 0.4 \citep{fer06, jan10}.
We divide these ellipses into four sectors azimuthally: northeast (NE), northwest(NW), southeast (SE), and southwest(SW) (Figure \ref{fig:map_sn}). We further subdivide these sectors into four regions along the major axis distance ($R_{\rm maj} < 10\arcmin$, $10\arcmin < R_{\rm maj} < 25\arcmin$, $25\arcmin < R_{\rm maj} < 65\arcmin$, and $R_{\rm maj} > 65\arcmin$). Figure \ref{fig:map_sn} shows the $S/N$ of the co-added spectra of the blue and red GCs in subregions, ranging from 6 to 68. We measured the mean ages and total metallicities for 16 and 14 co-added spectra for the blue and red GCs, respectively.

\begin{figure}
\includegraphics[width=\columnwidth]{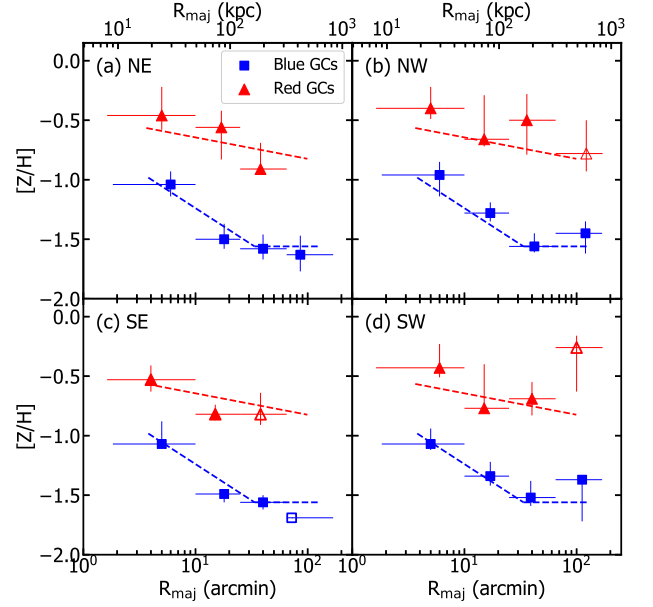}
\caption{[Z/H] vs. major axis distance from M87 for the blue GCs (squares) and red GCs (triangles) in four different sectors: (a) NE, (b) NW, (c) SE, and (d) SW.
The blue and red dashed lines are the same in all plots, and indicate the radial gradients in the metallicities of the full blue and red GC sub-populations, respectively (Figure \ref{fig:r_measure}(f)).
The filled and open symbols represent the GC samples with co-added spectral $S/N$ higher and lower than 30, respectively.
\label{fig:zhgrad_pa}}
	\end{figure}

The blue and red GCs in 26 out of 30 subregions are older than 10 Gyr, not showing specific trends. In addition to the sectors, we investigated the substructure that \citet{pow18} identified as the region that relatively young GCs populate, located at 11$\farcm$5 south of M87 with a diameter of $\sim$ 6$\arcmin$. We found 41 GCs in that region and measured their mean age with the co-added spectrum. They have a mean age of 9 Gyr, which is indeed a few Gyr smaller than the azimutally averaged age of other GCs at the same distance from M87. Further kinematic studies of the GCs in this region will be helpful to reveal the origin of any younger GC population.

The two GC populations show radial gradients in metallicity for each sector. Figure \ref{fig:zhgrad_pa} shows the radial trends in metallicity of the blue and red GCs in four sectors. The radial trends of the blue GCs in all sectors are consistent with that derived with the entire blue GC sample (Figure \ref{fig:r_measure}(e)). The red GCs also show shallow radial gradients in general, but the red GCs in the NW region of M87---the direction of M86 and M84---have slightly elevated mean metallicities compared to those in other sectors (Figure \ref{fig:zhgrad_pa}(b)). There is a possibility that this metal-rich population originates from massive galaxies in the northwest region of M87. However, the difference is not significant enough for us to draw a definite conclusion.


\section{Discussion: The Assembly History of M87 and the Virgo Cluster Core\label{section:discussion}}

We have estimated ages and metallicities from co-added spectra of GC sub-populations and have found several notable features, especially in GC metallicity variations as a function of projected distance from M87. The GCs in M87 and the intracluster region are mostly old, potentially allowing us to trace the mass assembly history of M87 to early epochs ($>$ 10 Gyr). In this section, we discuss the influence of two different parameters related to the assembly history of the Virgo core region and M87 itself: the masses of GC progenitor host galaxies and the dependence on environment.

\subsection{Gradients and GC Progenitor Galaxy Masses}

The chemical abundances of stars and star clusters are intimately tied to the mass of the galaxy within which they are formed. The stellar mass-metallicity relation of galaxies has connected these quantities since its discovery \citep{leq79}, and the relation was confirmed with gas-phase metallicities, especially for star-forming galaxies \citep[e.g.,][]{tre04, man10, am13}. A series of studies have also clarified and extended the relation with stellar metallicities for SDSS star-forming galaxies \citep{gal05} down to Local Group dwarf spheroidal and irregular galaxies \citep{kir13}. Although the metallicities of the GCs are about 0.8--1~dex lower than those of the stars in a galaxy \citep{jor04,lam17}, we expect that the metallicities of the GCs also represent masses of progenitors where the GCs were formed. Therefore, one interpretation of the metallicity radial gradients seen in the blue and red GCs is that both blue and red GCs in the inner region of M87 were formed in more massive galaxies than those in the outskirts. 

\begin{figure*}
\centering
\includegraphics[width=0.9\textwidth]{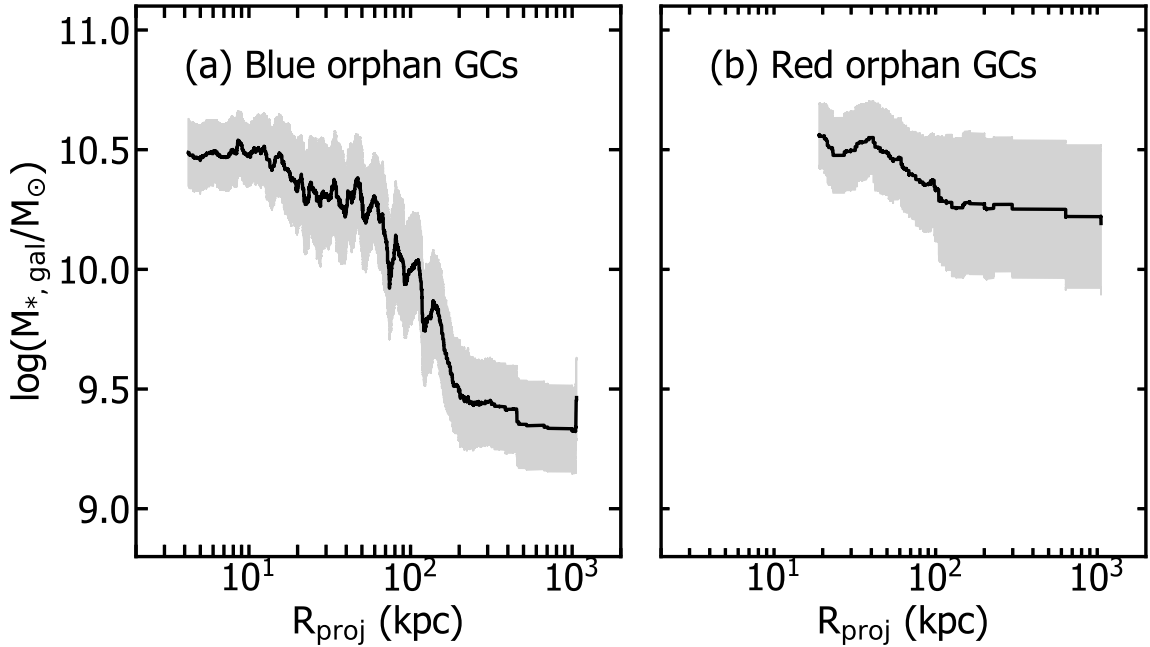}
\caption{(a) Mean stellar masses of progenitors at infall time as a function of projected clustercentric distance for blue orphan GCs in a simulated galaxy cluster \citep{ram20}. (b) Same as (a), but for red orphan GCs. The solid lines indicate the running averages of the progenitor masses with moving bins of $N = 100$. The shaded regions represent the standard deviation from the running averages.
Note that this plot does not include any GC populations formed {\it in-situ}.
\label{fig:minf_r}}
	\end{figure*}

If the GCs in massive galaxies like M87 largely originate from merged and accreted galaxies that were disrupted \citep[e.g.,][]{cot98}, and GC metallicity correlates with progenitor host mass, then a metallicity gradient suggests a radial gradient in the stellar masses of disrupted progenitors. This is a natural consequence of the efficiency of dynamical friction as a function of the merger ratio \citep{amo19}. Merged galaxies will spiral into the center due to dynamical friction until they are disrupted. More massive progenitors not only experience dynamical friction more efficiently, they can resist complete disruption until they are deeper in the potential well of the larger galaxy. In this way, one expects more metal-rich stars and star clusters to be deposited closer to the center of the accreting galaxy, creating a metallicity gradient. 

To check whether this explanation can work in practice, we use the Illustris cosmological simulation \citep{vog14a, vog14b}, where we analyze the GC system of a simulated galaxy cluster with a virial mass consistent with that of the Virgo Cluster as an example. \citet{ram20} tagged GCs to individual galaxies in the simulated galaxy at their infall time, following a Hernquist density profile \citep{her90}. The infall time here is defined as the epoch when a given galaxy can no longer be identified as the central galaxy of its own friends-of-friends group. They adopted different values of the characteristic density and scale length in the Hernquist profile for blue and red GC systems. The relation between the total mass in GCs and host galaxy mass \citep{hhh15, hud15} is taken into account to assign masses to GCs. The detailed GC tagging analysis is described in \citet{ram20}. In our analysis, we only study ``orphan" globular clusters---GCs for which the progenitor host galaxies are destroyed at the present day.

We investigate where orphan GCs are located as a function of their progenitors' masses.
The mean magnitude of our spectroscopic sample is $\langle g'\rangle= 21.5$~mag, with 99\% of the sample having $g'<22.5$~mag. The latter limit corresponds to a mass of $M = 4 \times 10^5 M_\odot$ with an assumption of a GC mass-to-light ratio in the $g$-band of 2. We therefore only select the simulated galaxies with GC system masses higher than $4 \times 10^5 M_\odot$ for a fair comparison. These galaxies have stellar masses ranging from $3\times10^8$ to $10^{11}~M_\odot$.
Assuming that the mass of a GC is $4\times 10^5 M_\odot$, we randomly chose GC particles until their total mass was equal to the total mass of GCs in each galaxy. We repeated this particle sampling 100 times to determine the mean trend of progenitors' masses as a function of projected distance from the central galaxy.


Figure~\ref{fig:minf_r} shows the mean stellar masses at infall time of the progenitor galaxies of orphan GCs as a function of their present day projected clustercentric distance. 
We find a steep gradient in the mean progenitor masses for blue orphan GCs located at 50 kpc $< R_\mathrm{proj} <$ 300 kpc, ranging from $10^{10.5} M_\odot$ to $10^{9.3} M_\odot$. This implies that the blue orphan GCs in the inner region of the central galaxy originate from more massive satellites that are more likely to sink further into the center than those in the outer region. We cannot constrain the gradient in the innermost region ($R_\mathrm{proj} <$ 50~kpc) because in this analysis, we do not include the ``{\it in-situ}" population that has formed at the center and would be the most metal-rich population in the central galaxy. On the other hand, the red orphan GCs do not show a steep gradient in the mean progenitor masses because the progenitor masses of the red GCs have much narrower range than those of the blue GCs. The red GCs are rarely formed in galaxies with $M_{\rm *, infall} < 10^{10} M_\odot$. This trend matches well with the observed radial gradient in metallicity for the red GCs being shallower than that of the blue GCs.

In addition, we also found that the orphan GCs, especially blue ones, from progenitors with the earliest infall time ($z \sim 3.6$) are located at the innermost region in this simulation, while those from recent infallers with a mean infall time of $z \sim 2.5$ are distributed in the outer region. This hints at the observed age difference of a few Gyr between the inner and outer GCs.

While the simulations have their limitations, this example illustrates how including GCs in cosmological galaxy simulations, in a way that allows for studies of GC properties as a function of their spatial distributions \citep[e.g.,][]{ram20,rei21,che22}, makes available new diagnostics for galaxy assembly history.

\subsection{GC Occupation Fraction and Gradient Flattening}

In addition to the steeper radial metallicity gradient for blue GCs, we have found that the gradient flattens beyond $\sim165$~kpc (Figure \ref{fig:r_measure}(f)). Since the mass-metallicity relation extends to quite low metallcities and masses, and the gravitational processes that govern dynamical friction and galaxy disruption are also continuous, it might seem surprising that we should find any break in these radial trends. We show below that this could be related to the GC occupation fraction---the fraction of galaxies that contain GCs---as a function of galaxy mass.

\citet{san19} presented the star cluster occupation fraction of Virgo early-type galaxies as a function of galaxy stellar mass (see their Figure~6). All galaxies with $M_{*}>10^{9}~M_{\odot}$ host GCs with $g'<24.5$~mag (roughly the mean of the GC luminosity function), while only half of the galaxies with $M_{*} \sim 10^{7}~M_{\odot}$ do.
Because of the flux limit of our survey, we only observed relatively bright GCs, with a mean sample magnitude of $\langle g'\rangle=21.5$~mag, and with 99\% of the sample having $g'<22.5$~mag. The relevant GC occupation fraction for determining whether an accreted progenitor contributes GCs to our observed sample is therefore whether it contains a GC with $g'<22.5$~mag. Using the GC data from the ACS Virgo Cluster Survey \citep{cot04,jor09}, we determine the GC occupation fraction for Virgo galaxies as a function of GC magnitude and galaxy stellar mass. For galaxies with $M_{*} \sim 10^{10}~M_{\odot}$, the occupation fraction for GCs with $g'<22.5$~mag is $\sim 80\%$, but for galaxies with $M_{*} \sim 10^{9}~M_{\odot}$, the occupation fraction drops to $\sim 25\%$. Galaxies with even lower stellar masses are then not expected to host many GCs with $g' < 22.5$ mag. Therefore, we expect that beyond $\sim$ 100-150 kpc, while many lower mass galaxies are being accreted and disrupted, only a small fraction of them can contribute GCs. {\it The flattening of the gradients at large distances is a natural by-product of the fact that, below a certain mass, the low-mass galaxies that would host even more metal-poor GCs have few to no GCs}. We see a sign of this in the simulated GC system shown in Figure~\ref{fig:minf_r}(a). Beyond a certain distance, the mean progenitor masses of blue orphan GCs become constant because in this simulated galaxy cluster, the galaxies disrupted in the outskirts have lower masses and do not host any GCs more massive than $4 \times 10^5 M_\odot$. The location of the break point likely depends on the mass and assembly history of individual galaxy clusters. While the flux limit of our survey may affect the exact [Z/H] or [$\alpha$/Fe] value at which the gradient flattens, we expect the general trend would be similar even if one could observe fainter GCs, and this could be tested with future observations.

We present this as an alternative to a more traditional explanation for stellar population gradients, which is that they occur during dissipational star formation, where more rapid star formation and gas recycling toward the galaxy center causes more rapid chemical evolution. Evidence for steep metallicity gradients in ``relic'' galaxies with little apparent accreted populations \citep{bea18,mar15} suggests that {\it in situ} star formation likely plays a role in these gradients, especially for more metal-rich populations. However, given the complexity of modeling a dissipational collapse, we find it interesting to show that the trends we see in the M87 GC system can be explained using a picture dominated by dry merging. This is especially relevant as it is expected that the intracluster light is dominated by stars acquired through dissipationless accretion \citep{pbz07}. It is likely however, that dissipational collapse plays some role in forming the gradients we observe, especially in the inner regions, and it will be interesting to explore that possibility with future simulations.

	\subsection{[$\alpha$/Fe] and Dependence on Environment}

\begin{figure}
\centering
\includegraphics[width=\columnwidth]{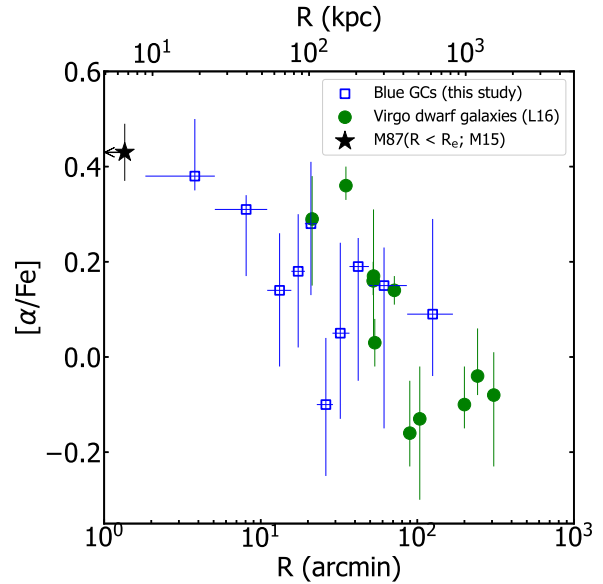}
\caption{[$\alpha$/Fe] values of blue GCs (blue squares) and Virgo dwarf galaxies \citep[green circles;][]{liu16} as a function of major axis distance from M87. \label{fig:afe_compare}}
	\end{figure}

The [$\alpha$/Fe] ratios allow us to understand the star formation timescales in the host environments.
There is a well-known relation between $\alpha$-element abundances and central velocity dispersion values for galaxies. The larger velocity dispersion the galaxies have, the higher [$\alpha$/Fe] values they have \citep{tho05,ann07,mcd15}. However, for low-mass galaxies, the local density could be a more critical parameter than the mass. \citet{liu16} measured the $\alpha$-element abundances of 11 low-mass dwarf galaxies in the Virgo Cluster and found the [$\alpha$/Fe] values of dwarf galaxies closer to M87 are higher than those distant from M87. 

Similarly, we find moderate negative radial gradients in mean [$\alpha$/Fe] values, especially for the blue GC sample, a population that likely originates from low-mass galaxies. This suggests that the blue GCs in the central region of M87 were formed in progenitors where rapid star formation occurred. Figure \ref{fig:afe_compare} shows the comparison of [$\alpha$/Fe] gradients of GCs from this study and dwarf galaxies from \citet{liu16}. The gradients are similar to each other although the GCs are more centrally concentrated (in projected radius) than the Virgo dwarfs.

Following the simulation result above, the blue GCs in the inner region of M87 might originate from relatively massive galaxies that became satellites of the central galaxy at high redshift. Intense star formation and rapid quenching occurred in these massive progenitors in the high-z Universe, resulting in high [$\alpha$/Fe] abundances in their stars and star clusters. Similarly, dwarf galaxies in dense environments have enhanced $\alpha$-element abundances \citep{chi08,smi09,liu16}. There is a possibility that the most metal-rich of the blue GCs in the inner region originate from dwarf galaxies that were quenched quickly by the dense environment \citep{mis16}.

\subsection{Local substructures}

The Virgo Cluster is a dynamically young galaxy cluster with several subclusters \citep{bst85, bts87, bpt93}. \citet{mih05, mih17} presented the deep imaging survey showing the diffuse intracluster light in the Virgo core region, indicating the ongoing interaction between galaxies. Related to this dynamical status, it is expected that the GCs can be stripped not only from low-mass dwarf galaxies but also from massive galaxies. The observed metal-rich GCs imply the ongoing interaction between M87 and nearby massive galaxies.

We detected a hint that the metal-rich GCs located in the NW region of M87 (Figure \ref{fig:zhgrad_pa}) are slightly more metal-rich than in other areas of the Virgo core. This direction is toward massive early-type galaxies, M86 and M84. We suspect that these metal-rich GCs might be associated with M84 because we exclude the GCs that have lower radial velocities corresponding to the systemic velocity of M86 (Figure \ref{fig:subpop}(a)). We defer to a future paper an investigation of the many GCs associated with known substructures, particularly the M86 group. 

\section{Summary \label{section:summary}}
 
We present a stellar population analysis of the GCs in the central region of the Virgo Cluster, especially around M87, using the MMT/Hectospec spectra. We select 692 GCs that belong to the main body of the Virgo cluster to investigate the stellar populations of GCs, excluding the GCs associated with substructures. The GCs are divided into several groups with various criteria for colors, major axis distances, and position angles. We measured mean ages, total metallicities, [Fe/H], and $\alpha$-element abundances of the GC sub-populations using their co-added spectra based on Lick indices. The main findings of this study are summarized as follows.

\begin{itemize}

\item The GCs as a whole follow a two-component radial metallicity ([Z/H]) trend, with a steep gradient following a power-law slope of $-0.84^{+0.15}_{-0.11}$ within $R_{\rm maj} = 22\arcmin$ (109~kpc), and a flat gradient with a mean metallicity of ${\rm [Z/H]} \sim -1.34$ beyond. The trends for [Fe/H] and [$\alpha$/Fe] are similar, with inner region slopes of $-0.52^{+0.06}_{-0.14}$ and $-0.34\pm0.11$, and outer flattenings at ${\rm [Fe/H]} = -1.48^{+0.03}_{-0.07}$ and ${\rm [\alpha/Fe]} = 0.15\pm0.04$, respectively.

\item When analyzing gradients in GC color sub-populations (blue and red), we find that the two-component trend is clearly seen only in the blue GC system, while the red GCs are best fit by a single component. For [Z/H], the blue GCs exhibit an inner region power-law slope of $-0.61^{+0.05}_{-0.21}$ within $R_{\rm maj} = 34\arcmin$ (165~kpc) and a mean metallicity of ${\rm [Z/H]} = -1.56\pm0.05$ beyond. The red GCs show a much shallower radial gradient in [Z/H] than do the blue GCs, with a power-law slope of $-0.18^{+0.09}_{-0.15}$, and no obvious flattening at large radius. 

\item The [Fe/H] and [$\alpha$/Fe] gradients also behave differently for the blue and red GC sub-populations. The blue GCs follow the same two-component trend, with inner region slopes of $-0.32^{+0.16}_{-0.10}$ and $-0.30^{+0.08}_{-0.17}$, and outer flattenings at ${\rm [Fe/H]} = -1.64^{+0.09}_{-0.05}$ and ${\rm [\alpha/Fe]} = +0.09^{+0.04}_{-0.10}$, respectively. These results show that the [Z/H] gradient seen in the blue GCs is driven in roughly equal portions by [Fe/H] and [$\alpha$/Fe]. By contrast, the red GCs exhibit only a weak gradient in [Fe/H] (slope of $-0.16^{+0.15}_{-0.05}$), and no gradient in [$\alpha$/Fe]. 

\item We investigated the azimuthal variation of GC metallicities and detected a hint of local substructures toward the massive early-type galaxy M84, consisting of metal-rich GCs.

\item The GCs in the Virgo core are mostly old ($>$ 8 Gyr) regardless of their colors and locations. The outer GCs are formally about 2 Gyr younger than inner GCs, but there is a possibility that the presence of blue horizontal branch stars are causing this apparent age gradient.

\end{itemize}

We interpreted the radial gradient of GCs in metallicity and [$\alpha$/Fe] in the context of the assembly history of M87 and the Virgo Cluster core. We used the Illustris cosmological simulations to investigate masses of disrupted GC progenitors in a simulated cluster.

\begin{itemize}
\item According to the mass-metallicity relation of galaxies, the more metal-rich blue GCs at smaller galactocentric distances were probably formed in more massive galaxies. We expect that these more massive progenitors were better able to survive disruption as they approached the galaxy center due to dynamical friction.
From the simulation, we found that the GCs from massive progenitors sink further into the central galaxy, while GCs from lower-mass progenitors are located preferentially at larger distances.

\item The flattening of metallicity gradients of blue GCs at large distances ($R \gtrsim 100$ kpc) is a natural consequence if the lower-mass dwarfs that could contribute GCs with even lower metallicity actually host few or no GCs.


\item We suggested that the dense environment around M87 could result in the high [$\alpha$/Fe] values of the GCs in the central region, mirroring the environmental dependence of the [$\alpha$/Fe] of Virgo dwarf galaxies \citep{liu16}.

\end{itemize}

Overall, we suggest a scenario where more massive progenitors accreted into the central galaxy (M87) contribute a significant fraction of the GCs in the inner region.
Future simulation studies on various galaxy clusters will be helpful to understand the relation between the assembly history of a host galaxy cluster and chemical properties of the GCs residing in it.


\acknowledgments
Observations reported here were obtained at the MMT Observatory, a joint facility of the University of Arizona and the Smithsonian Institution. MMT telescope time was granted, in part, by NOAO, through the Telescope System Instrumentation Program (TSIP). TSIP is funded by NSF. EWP thanks the staff of the MMT Observatory for their professional support over multiple observing runs. EWP also thanks the Harvard-Smithsonian Center for Astrophysics and its Institute for Theory and Computation for hosting him during an unexpectedly long pandemic visit.
We thank Daniel Fabricant for sharing his flux calibration routine for MMT/Hectospec data. We also thank the anonymous referee for useful comments. 

This paper is based on observations obtained with MegaPrime/MegaCam, a joint project of CFHT and CEA/IRFU, at the Canada-France-Hawaii Telescope (CFHT), which is operated by the National Research Council (NRC) of Canada, the Institut National des Sciences de l’Univers of the Centre National de la Recherche Scientifique (CNRS) of France, and the University of Hawaii. This research used the facilities of the Canadian Astronomy Data Centre operated by the National Research Council of Canada with the support of the Canadian Space Agency. 

CL acknowledges support from the National Natural Science Foundation of China (NSFC, Grant No. 12173025, 11673017, 11833005, 11933003), 111 project (No. B20019), and Key Laboratory for Particle Physics, Astrophysics and Cosmology, Ministry of Education.
AL is supported by Fondazione Cariplo, grant No 2018-2329.
AL, THP, and PAD acknowledge support from grant ANR-19-CE31-0022(POPSYCLE) of the French Agence Nationale de la Recherche.
AJ and SE acknowledge support from ANID -- Millennium  Science  Initiative -- ICN12\_009.
MGL was supported by the National Research Foundation grant funded by the Korean Government (NRF-2019R1A2C2084019).
H.S.H. acknowledges the support by the National Research Foundation of Korea (NRF) grant funded by the Korea government (MSIT) (No. 2021R1A2C1094577).
S.L. acknowledges the support from the Sejong Science Fellowship Program through the National Research Foundation of Korea (NRF-2021R1C1C2006790).


\clearpage

\listofchanges

\end{document}